\documentclass[fleqn,10pt]{wlscirep}
\usepackage[utf8]{inputenc}
\usepackage[T1]{fontenc}
\usepackage{blindtext}
\usepackage{gensymb}
\usepackage{graphicx}
\usepackage[euler]{textgreek}
\usepackage{braket}
\usepackage{physics,amsmath}
\usepackage{cleveref}

\title{Magnetic sensitivity enhancement via polarimetric excitation and detection of an ensemble of NV centers}
\author[1]{Simone Magaletti}
\author[1]{Ludovic Mayer}
\author[1]{Xuan Phuc Le}
\author[1,*]{Thierry Debuisschert}
\affil[1]{Thales Research and Technology, 91767 Palaiseau Cedex, France.}
\affil[*]{thierry.debuisschert@thalesgroup.com}

\begin{abstract}
The negatively charged nitrogen-vacancy center (NV) presents remarkable spin-dependent optical properties that make it an interesting tool for magnetic field sensing. In this paper we exploit the polarization properties of the NV center absorption and emission processes to improve the magnetic sensitivity of an ensemble of NV centers. By simply equipping the experimental set-up of a half-wave plate in the excitation path and a polarizer in the detection path we demonstrate an improvement larger than a factor of two on the NV center magnetic sensitivity.
\end{abstract}
\begin{document}
\flushbottom
\maketitle
%
%
\thispagestyle{empty}
\section*{Introduction}
\begin{figure}
	\centering
	\includegraphics[width=0.7\linewidth]{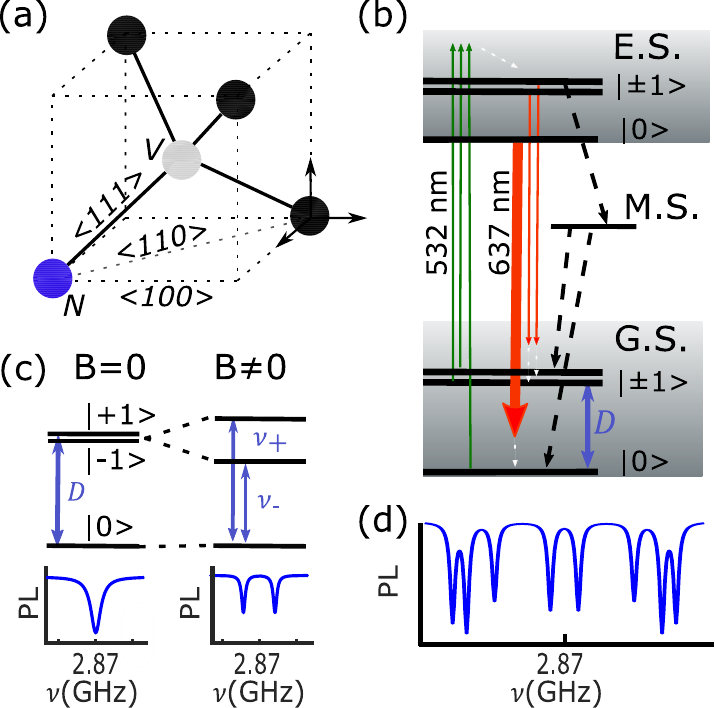}
	\caption{NV center properties.
		(a) Nitrogen-Vacancy center hosted in the diamond lattice. Nitrogen is represented by a blue sphere, carbon vacancy by a white sphere, carbon atoms by black spheres. The $\langle100\rangle$, $\langle110\rangle$ and $\langle111\rangle$ crystallographic directions are also indicated.
		(b) NV center energy levels. Green arrows represent the non-resonant optical pumping from the ground state (G.S.) to the excited state (E.S.). Red arrows represent the radiative transitions; the width of the arrow is qualitatively related to the PL emission rate. Black dashed arrows represents the ISC process through the metastable state (M.S). The transition from the E.S.$\ket{0}$ spin sublevel to the M.S. is 10 times less probable than the one from $\ket{\pm1}$ and it is not represented \cite{Robledo2011}. D=2.87~GHz is the ground state zero field splitting. (c) NV center ground state spin sublevel without (left) and with (right) an external static magnetic field applied. The ODMR spectrum is also represented. $\nu_\pm=|D\pm\gamma B_{NV}|$ are the resonance frequencies of the $\ket{0}\rightarrow\ket{\pm1}$ transitions. (d) ODMR spectrum of an ensemble of NV centers. Eight peaks, two for each of the four NV center families, are visible.}
	\label{fig:fig2}
\end{figure}

The negatively charged nitrogen-vacancy center (NV) is a spin-1 color center in diamond composed by a nitrogen atom and a carbon vacancy in two adjacent positions of the lattice (\cref{fig:fig2}a).
It presents remarkable spin-dependent optical properties and a long spin-coherence time (ms) at room-temperature that allow for optical polarization, optical read-out and coherent manipulation of its electron spin \cite{Doherty2013}. 
These features make it an appealing tool for quantum technologies and in particular for magnetic field sensing \cite{Rondin2014}, where magnetic field sensitivity as low as a few pT$\cdot$Hz$^{-1/2}$ have been demonstrated \cite{Zhang2021,Wolf2015,Wang2022}. The NV center magnetic sensitivity \texteta~scales as \cite{Barry2020}:
\begin{equation}
\eta\propto\frac{\Delta\nu}{C\sqrt{S_0}}
\label{eq:SensitivityScaling}
\end{equation}
where (\textDelta\textnu) is the linewidth of the spin resonance \cite{Bauch2018}, (S\textsubscript{0}) is the detected photoluminescence (PL) and (C) is the contrast of the measurement, namely the relative photoluminescence variation between the system on resonance and out of resonance. Since the sensitivity scales with the square root of S\textsubscript{0}, magnetic field measurements that do not require a nanometer-scale resolution are usually performed with an ensemble of NV centers.
Moreover, ensembles make possible the vector measurement of the static magnetic field \cite{Steinert2010,Chipaux2015}. 
Indeed, in the tetrahedral diamond lattice, four possible nitrogen-vacancy orientations (NV center axis) exist, each of them defining a so called NV center family (\cref{fig:fig2}a). In the low-magnetic-field regime (less than some tens of mT), each family provides information on the static magnetic field component along its specific axis. In this way, the system is inherently a vector magnetometer \cite{Steinert2010,Chipaux2015,Levine2019}. 
However, when measuring the magnetic field component along one diamond crystallographic axis, the NV centers oriented along other directions act as sources of noise since they emit a PL background that reduces the contrast of the measurement (\cref{eq:SensitivityScaling}). This effect is even more adverse for applications that exploit only a single NV center family \cite{Soshenko2021,Horsley2018,Magaletti2022}. The straightest way to suppress the PL background emitted by the NV center families that do not participate to the measurement is by working with diamond samples that host NV centers preferentially oriented along only one (or some) diamond crystallographic direction \cite{Miyazaki2014}. The classical growth along a $\langle100\rangle$ direction does not lead to a preferential orientation, but growing diamonds along $\langle110\rangle$, $\langle113\rangle$ or $\langle111\rangle$ crystallographic directions allows reaching 50\% ($\langle110\rangle$ \cite{Edmonds2012}), 73\% ($\langle113\rangle$ \cite{Lesik2015}) and 99\% ($\langle111\rangle$ \cite{Fukui2014}) of preferential orientation of NV centers. However, in diamond crystals grown along $\langle111\rangle$ direction, it is difficult to achieve high NV concentrations and low levels of parasitic impurities. Moreover, standard post-treatments, such as irradiation and annealing, used to increase the NV center concentration tend to deteriorate the as-grown preferential orientation, thus limiting the gain in magnetic sensitivity\cite{osterkamp2019}. In addition, samples showing preferential orientations are not easily exploitable as vector magnetometers.
Another technique to increase the contrast of one family relies on applying to the other ones strong transverse magnetic field components to induce a photoluminescence quenching \cite{Tetienne2012}. This method, however, imposes constraints on the external static magnetic field thus limiting its applicability.

In this paper, we propose a different approach that exploits the polarization properties of the NV center emission and absorption processes to tune respectively the NV center excitation probability and the PL collection efficiency. Previous works investigated these two contributions and employed them principally to identify the orientation of NV centers \cite{Christinck2020,Alegre2007,Bayer2020,Backlund2017,Reuschel2022}. Here, we exploit both of them to tune and optimize the contrast and the magnetic sensitivity (\cref{eq:SensitivityScaling}) of one NV center family with respect to the others. In particular, we propose and realize an easy-to-implement configuration that allows doubling the magnetic sensitivity simply by equipping the experimental set-up of a half-wave plate in the excitation path and a polarizer in the detection path.

\section*{The Nitrogen-Vacancy center}
The energy level structure of the NV center is depicted in \cref{fig:fig2}b. The ground state and the exited state are spin triplets while the metastable state is a spin singlet. Spin sublevels are defined as $\ket{m_s}$ states, where m\textsubscript{s} is the spin projection quantum number along the NV center axis. 
In absence of any static magnetic field, both the ground state and the exited state $\ket{\pm1}$ spin sublevels are degenerate and the ground state zero-field splitting D is equal to 2.87~GHz. At room temperature, the three spin sublevels are equally populated.
Under optical pumping \cite{Robledo2011}, two decay paths exist. First, a radiative spin-conserving decay, which is responsible of the NV center PL.
The emission spectrum (red arrows) has a zero-phonon line at 637 nm and a broad electron-phonon band (white dashed arrows), up to 800 nm. The radiative process is more probable when the NV center is in the $\ket{0}$ state than in the $\ket{\pm1}$ states, which makes possible the optical readout of the NV center spin state by monitoring its PL intensity level. Second, a non-radiative intersystem crossing process (black dashed arrows) which is more probable when the NV center is in the $\ket{\pm1}$ states than in the $\ket{0}$ state, and allows, after some optical cycles, the polarization of the system in the $\ket{0}$ state.

The two ground state spin resonances ($\ket{0}\rightarrow\ket{\pm1}$) allowed by the spin selection rules (\textDelta m\textsubscript{s}~=~$\pm$1) can be optically detected (ODMR) by sweeping the frequency of a radio frequency (RF) field while the NV center photoluminescence is monitored. The drop of PL indicates the spin resonance (\cref{fig:fig2}c). Applying a static magnetic field, the Zeeman interaction lifts the degeneracy between the $\ket{\pm1}$ states and thus two peaks are visible on the ODMR spectrum. At first order, the frequency difference (\textDelta f) between the two peaks depends on the magnetic field component (B\textsubscript{NV}) along the NV center axis:
\begin{equation}
\Delta f=2\gamma B_{NV}
\end{equation}
where the NV center gyromagnetic ratio \textgamma=28~GHz$\cdot$T$^{-1}$. When working with an ensemble of NV centers submitted to a small static magnetic field arbitrarily oriented, the ODMR spectrum usually consists of eight peaks, two for each of the four NV center possible orientations (\cref{fig:fig2}d). In this case, we define the ODMR contrast C\textsubscript{i} of the NV center family i as:
\begin{equation}
C_i=\frac{\text{PL}_i^{\text{OFF}}-PL_i^{\text{ON}}}{\sum_{j=1}^4 \text{PL}_j^{\text{OFF}}}
\label{eq:GenerzlContrast}
\end{equation}

Where PL\textsubscript{j}\textsuperscript{ON} and PL\textsubscript{j}\textsuperscript{OFF} are respectively the PL of NV centers belonging to the family j when they are on resonance and out of resonance with the RF field. 
\Cref{eq:GenerzlContrast} clearly points out the problem we deal with in this paper: the contrast reduction of one NV center family due to the PL background emitted by the others.
As discussed in the introduction, we address this point by studying the polarization properties of both the NV center excitation and its PL emission. These two processes are governed by two perpendicular identical dipoles laying in the plane orthogonal to the NV center axis \cite{Hossain2008}. As a consequence, the absorption process depends on the laser polarization and the emitted PL is polarized. At room temperature, due to the dynamic Jahn-Teller effect, the two dipoles equally emit independently from the one which is excited \cite{Fu2009,Kaiser2009}. Therefore, the PL polarization properties can be studied independently of the polarization of the laser field.

The excitation probability (P(\textbf{n\textsubscript{L}})) of a NV center oriented along \textbf{n\textsubscript{NV}} ($\|\textbf{n\textsubscript{NV}}\|=1$) by a laser whose electric field is oriented along \textbf{n\textsubscript{L}} ($\|\textbf{n\textsubscript{L}}\|=1$) reads as\cite{Dolan2014} (supplementary note 1):
\begin{equation}
P(\vectorbold{n_{L}})\propto 1-|\vectorbold{n_{NV}\cdot n_{L}}|^2
\label{eq:ExcitationProbability}
\end{equation}
Similarly, the PL intensity (I(\textbf{n\textsubscript{P}})) emitted along \textbf{u} ($\|\textbf{u}\|=1$ ) and polarized along \textbf{n\textsubscript{P}} ($\|\textbf{n\textsubscript{P}}\|=1$), with \textbf{n\textsubscript{P}} orthogonal to \textbf{u}, reads as (Methods\ref{App:B}, supplementary note 1):
\begin{equation}
I(\vectorbold{n_{P}})\propto 1-|\vectorbold{n_{NV}\cdot n_{P}}|^2
\label{eq:PLBeta}
\end{equation}
From an experimental point of view, \textbf{n\textsubscript{P}} corresponds to the axis of a polarizer in the PL detection system. The requirement of \textbf{n\textsubscript{P}} orthogonal to \textbf{u} corresponds to the approximation of collecting only photons propagating along the optical axis of the PL detection system (\textbf{u}) (Methods\ref{App:B}).

Moving to an ensemble of NV centers with no preferential orientation, according to \cref{eq:ExcitationProbability,eq:PLBeta}, the detected PL (S\textsubscript{0} in \cref{eq:GenerzlContrast}) when the NV centers are out of resonance and the ODMR contrast of the NV center family i (C\textsubscript{i} in \cref{eq:GenerzlContrast}) read as:
\begin{equation}
S_0(\vectorbold{n_{P}},\vectorbold{n_{L}})=\alpha\frac{N\cdot \text{PL}_0^{\text{OFF}}}{4}\sum_{j=1}^4 P_j(\vectorbold{n_{L}}) I_j(\vectorbold{n_{P}})
\label{eq:S0Polarization}
\end{equation}
 \begin{equation}
 C_i(\vectorbold{n_{P}},\vectorbold{n_{L}})=\frac{P_i(\vectorbold{n_{L}}) I_i(\vectorbold{n_{P}}) (\text{PL}_0^{\text{OFF}}-\text{PL}_0^{\text{ON}})}{\sum_{j=1}^4 P_j(\vectorbold{n_{L}}) I_j(\vectorbold{n_{P}}) \text{PL}_0^{\text{OFF}}}
 \label{eq:ContrastPolarization}
 \end{equation}
where N is the number of NV centers in the ensemble; \textalpha~is the PL collection efficiency of the imaging system; PL\textsubscript{0}\textsuperscript{ON/OFF} are the NV center intrinsic PL emission rates when it is on resonance and out of resonance with the RF field \cite{Robledo2011}. Although PL\textsubscript{0}\textsuperscript{ON/OFF} depend on the applied static magnetic field \cite{Tetienne2012}, for low static magnetic field we can approximate that they are independent of the NV center orientation.
We define the relative contribution (R\textsubscript{i}(\textbf{n\textsubscript{L},\textbf{n\textsubscript{P}}})) of the NV center family i to S\textsubscript{0} as:
\begin{equation}
R_i(\vectorbold{n_{P}},\vectorbold{n_{L}})=\frac{P_i(\vectorbold{n_{L}}) I_i(\vectorbold{n_{P}})}{\sum_{j=1}^4 P_j(\vectorbold{n_{L}}) I_j(\vectorbold{n_{P}})}=\frac{C_i(\vectorbold{n_{P}},\vectorbold{n_{L}})}{\sum_{j=1}^4 C_j(\vectorbold{n_{P}},\vectorbold{n_{L}})}
\label{eq:Rth}
\end{equation}
The last equality of \cref{eq:Rth} allows experimentally quantifying the contribution of each family to the total PL by measuring their ODMR contrast.

The NV center magnetic sensitivity (\cref{eq:SensitivityScaling}) depends on the laser polarization and the orientation of the polarizer axis through both the detected PL (S\textsubscript{0}) and the ODMR contrast (C\textsubscript{i}). We summarize this dependence in the parameter \textchi\textsubscript{i}, defined as:     
\begin{equation}
\chi_i(\vectorbold{n_{P}},\vectorbold{n_{L}})=R_i(\vectorbold{n_{P}},\vectorbold{n_{L}})\sqrt{\sum_{j=1}^4 P_j(\vectorbold{n_{L}})I_j(\vectorbold{n_{P}})}\propto C_i(\vectorbold{n_{P}},\vectorbold{n_{L}})\sqrt{S_0(\vectorbold{n_{P}},\vectorbold{n_{L}})}\propto\frac{1}{\eta_i(\vectorbold{n_{P}},\vectorbold{n_{L}})}
\label{eq:EtaPolarization}
\end{equation}
  
 When the four NV center families equally contribute to the detected PL, namely the laser polarization and the polarizer axis are oriented along a $\langle100\rangle$ direction and no preferential orientation is present, it results in R\textsubscript{i}=1/4 and \textchi\textsubscript{i}=0.33 (Methods\ref{App:C}). By numerically  maximizing \cref{eq:ContrastPolarization} and \cref{eq:EtaPolarization}, it is possible to find several pairs of (\textbf{n\textsubscript{L},\textbf{n\textsubscript{P}}}) that maximize the relative ODMR contrast or that optimize the sensitivity, by maximizing  \textchi\textsubscript{i}, of a given NV center family. The maximum value achievable for R\textsubscript{i} is 0.63 and for \textchi\textsubscript{i} is 0.79, which are respectively 2.5 and 2.4 times higher than what it is expected when the four NV center families equally contribute to the detected PL.

\section*{Results}
 \begin{figure}
	\centering
	\includegraphics[width=0.7\linewidth]{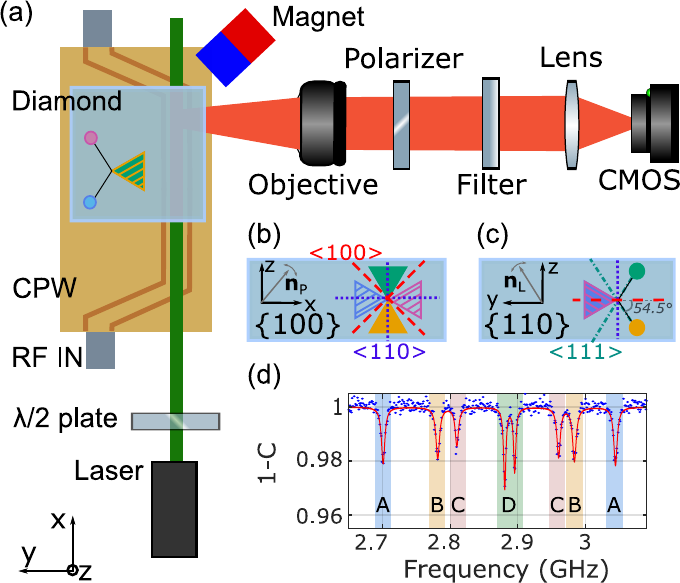}
	\caption{Experimental set-up. (a) The linearly polarized laser enters the diamond sample through a \{110\} plane. A half-wave (\textlambda~/2) plate allows turning its polarization. The PL is collected from the lateral \{100\} facet by means of an imaging system composed of a microscope objective, a polarizer, a filter, a lens and a camera. The RF field is brought in proximity of NV centers by means of a coplanar waveguide (CPW) on the top of which the diamond sample is glued. The orientation of the NV center families with respect to the diamond \{110\} top face are represented with the help of wedge-dash diagrams. Solid lines represent in-plane bonds; dashed lines represent bonds pointing out of the plane away from the viewer; wedge-shaped lines represent bonds pointing out of the plane toward the viewer. Line colors identify NV center families. (b) NV centers orientations with respect to the \{100\} facet from which the PL is collected. $\langle$100$\rangle$ and $\langle$110$\rangle$ diamond crystallographic directions are also represented by, respectively, dashed red lines and dotted blue lines. (c)  NV center orientations with respect to the \{110\} facet from which the laser excitation is performed. $\langle$100$\rangle$ and $\langle$110$\rangle$ diamond crystallographic directions are indicated using the same legend as (b). $\langle$111$\rangle$ directions are indicated by teal dash-dotted lines. (d) Normalized ODMR spectrum of the NV center ensemble under analysis. The fit (red line) is performed using \cref{eq:fitfunction}. The identification between the NV center families and the ODMR peaks is detailed in Methods\ref{App:E}.}
	\label{fig:fig4}
\end{figure}

We experimentally investigate the behaviour of an ensemble of NV center when both the laser polarization and the polarizer axis are tuned, and we show the opportunity of improving the ODMR contrast and the NV center magnetic sensitivity by simply equipping the experimental set-up of a half-wave plate in the excitation path and a polarizer in the detection path.
\begin{figure*}
	\centering
	\includegraphics[width=\textwidth]{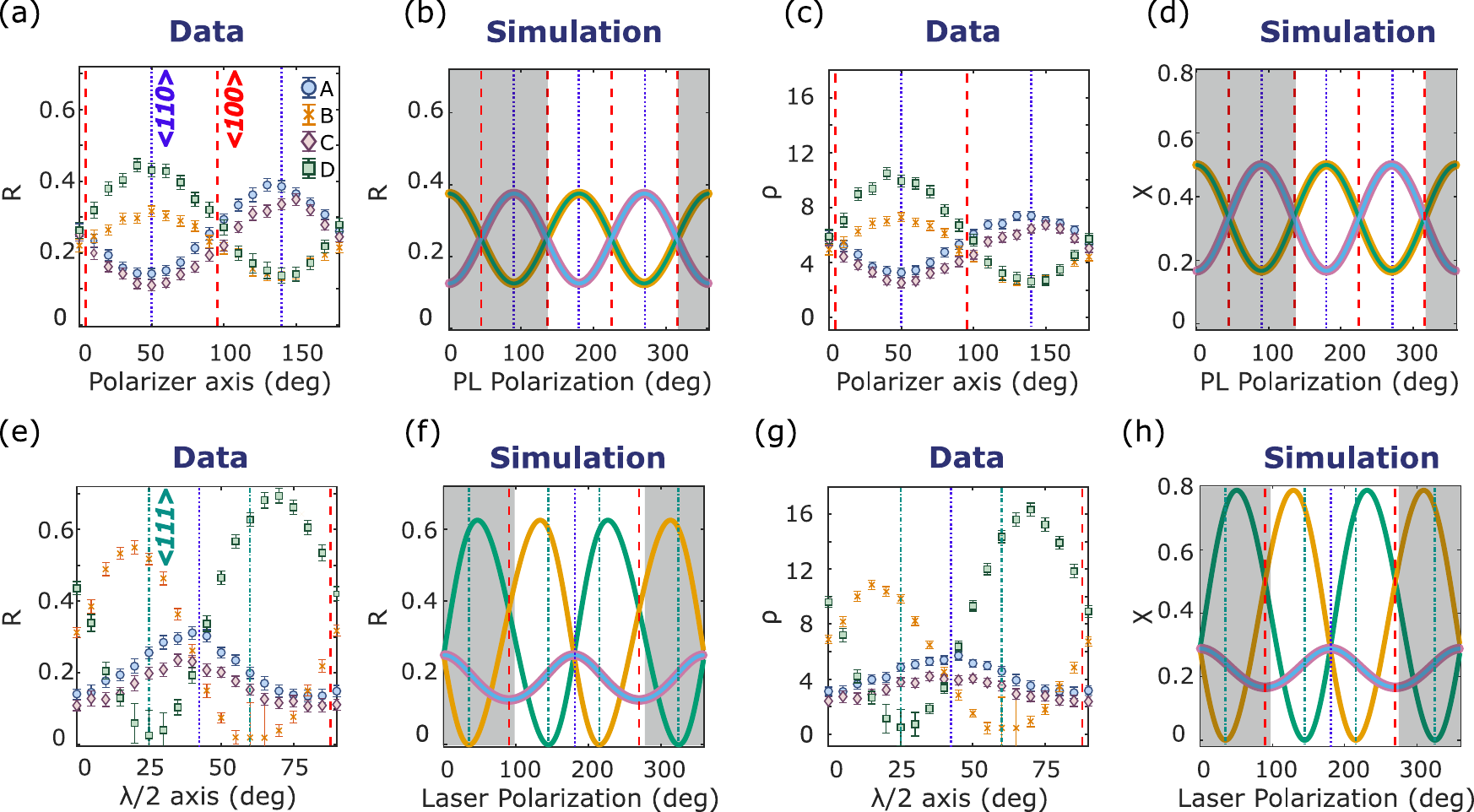}
	\caption{ODRM contrast and sensitivity dependence on polarizer axis and laser polarization. (a-d) Measurement (a) and simulation (b) of the relative contrast, \textrho~(c) and \textchi~(d) parameters of the four NV center families for different angles of the polarizer. Error bars result from the error propagation of the fit errors (estimated considering a 95\% confidence interval) of the contrast of each of the eight ODMR peaks.    
    According to \cref{fig:fig4}, A and D are the families whose $\ket{0}\rightarrow\ket{-1}$ transition resonates respectively at the lower and higher frequency. For the experimental results, the zero of the polarizer axis is arbitrarily chosen. For the simulation, it corresponds to the x axis of the laboratory reference frame (\cref{fig:fig4}b). Shadow areas are added a posteriori to match the measurement with the simulation.
    (e-h) Measurement (e) and simulation (f) of the relative contrast, \textrho~(g) and \textchi~(h) parameters of the four NV center families for different laser polarizations. In the simulation the zero of the laser polarization corresponds to the z-axis of the laboratory frame (\cref{fig:fig4}c). In Data the zero of the fast axis of the half-wave plate is set, with an accuracy of some degree, along the y axis of the laboratory frame (\cref{fig:fig4}a). Shadow areas are added a posteriori to match the measurement with the simulation. Error bars are defined as in (a) and (c). The large error bar for values of R and \textchi~close to zero is due to difficulties in doing the fit when the contrast is lower than the signal-to-noise ratio of the measurements. The plot is cut for values smaller than zero since they have no physical meaning.}
	\label{fig:fig5}
\end{figure*}
The experimental set-up is depicted in \cref{fig:fig4}. The sample is an optical grade CVD diamond crystal (Element Six) with two main \{110\} faces, as well as two \{110\} and two \{100\} lateral facets. A 532~nm linearly polarized laser, whose polarization is controlled by a half-wave plate, excites the NV centers through a \{110\} plane. The PL is collected from a \{100\} lateral facet by a microscope objective  (10X, NA 0.28), is spectrally filtered (Semrock FF01-695/75-25D) to suppress the background noise and is detected by a CMOS camera (ANDOR ZYLA-5.5). A polarizer selects the polarization of the detected PL.  
A neodymium magnet generates a static magnetic field so as to lift both the degeneracy between the $\ket{0}\rightarrow\ket{\pm1}$ transitions and to induce different resonance frequencies for each NV center family. The spin transitions are driven by a RF field brought in proximity of the diamond plate through a coplanar waveguide (CPW). The RF magnetic field is linearly polarized along a diamond $\langle$100$\rangle$ direction to equally excite the four NV center families.

The experiment consists in two steps. First, we set the laser polarization along a $\langle$100$\rangle$ diamond crystallographic direction (y axis of the laboratory reference frame, \cref{fig:fig4}a) to equally excite the four NV center families, and we measure the NV centers ODMR spectra tuning the polarizer angle in a \{100\} plane (\cref{fig:fig4}b). Second, we fix the polarizer axis and we measure the ODMR spectra tuning the laser polarization in a \{110\} plane through a half-wave plate (\cref{fig:fig4}c).
For each ODMR spectrum (\cref{fig:fig4}d), we evaluate the contrast of the eight ODMR peaks (Methods\ref{App:D}) and we define the contrast of the generic NV center family i (C\textsubscript{i}) as the mean value of the ODMR contrast of its two spin transitions. In this way, using \cref{eq:Rth}, we can measure the relative contribution (R\textsubscript{i}) of each NV center family to the total PL.
Similarly, according to \cref{eq:EtaPolarization}, we define the parameter \textrho\textsubscript{i}, which account for the sensitivity dependence on the laser polarization and the polarizer orientation in the experiment, as:
\begin{equation}
\label{eq:etaSperimental}
\rho_i= C_i\sqrt{S_{0}}\propto\chi_i
\end{equation}
where S\textsubscript{0} is the PL measured when NV centers are out of resonance with the MW field.
We named the NV center families A, B, C, D being A and D the families whose $\ket{0}\rightarrow\ket{-1}$ transitions resonate respectively at the lowest and the highest frequency (\cref{fig:fig4}d, \cref{fig:fig5}a). The identification between ODMR peaks and NV center families is described in Methods\ref{App:E}.

The relative contrast of each NV center family for different angles of the polarizer axis is reported in \cref{fig:fig5}a and shows a good agreement with the simulation realized using \cref{eq:Rth} and plotted in \cref{fig:fig5}b. The larger discrepancy from simulations of family D than other families is mainly attributed to spin state mixing induced by the transverse component of the static magnetic field \cite{Moreva2020}, which is stronger for this family than the others (\cref{fig:fig4}d). Families A-C and B-D have similar behaviours because of their symmetry with respect to the \{100\} plane from which the PL is collected. The maximum relative contrast (data, mean value over the four families: 0.37~$\pm$~0.06; simulation: 0.375) is achieved when the polarizer axis is perpendicular to the NV axis. The minimum (data: 0.13~$\pm$~0.02; simulation: 0.125) when it is along the projection of the NV center axis on the \{100\} plane. Positions of maxima and minima allow identifying the diamond $\langle$110$\rangle$ axis laying on the \{100\} plane from which the PL is detected.   
The sensitivity, here investigated through the \textrho\textsubscript{i} (\cref{fig:fig5}c) and \textchi\textsubscript{i} (\cref{fig:fig5}d) parameters, behaves similarly to the relative contrast. This happens for two reasons. First, the sensitivity scales linearly with the contrast and only as the square root of the detected PL (\cref{eq:EtaPolarization,eq:etaSperimental}). Second, the contrast depends more on the angle of the polarizer than the detected PL does (Methods\ref{App:F}, see supplementary note 1).

In the second part of the experiment, ODMR spectra are acquired for different laser polarizations. The polarizer is set so as to maximize the ODMR contrast of families B and D in \cref{fig:fig5}a. The measurement and the simulation of the relative contrast of each NV center family are respectively plotted in \cref{fig:fig5}e-f and they show a good agreement. Rotating the laser polarization in a \{110\} plane (\cref{fig:fig4}c) it is possible to align the electric field with one of the two NV center families laying in that plane ($\langle$111$\rangle$ axis), not exciting it anymore \cref{eq:ExcitationProbability}, and thus suppressing its contribution to the total PL (family B for \textlambda/2 axis angle of 60$\degree$, family D for \textlambda/2 axis angle of 25$\degree$ in \cref{fig:fig5}e). For these two families, it is interesting to observe that while the minimum of the contrast is achieved when the laser is aligned along their axis, the maximum is not achieved when the laser is perpendicular to it. For instance, looking at the simulation of the relative contrast of family B (\cref{fig:fig5}f), the minimum is achieved for a laser polarization angle of 35$\degree$ while the maximum at 133$\degree$ (instead of 125$\degree$). This is due, and it is the core idea of this paper, to the fact that the maximization of the contrast of one family is achieved not only increasing the contribution of that family to the total PL, but also minimizing the contribution of the others (see supplementary note 1).     
Looking at \cref{fig:fig5}f, the maximum ODMR relative contrast for family B is (R\textsubscript{B}(20$\degree$)=0.55$\pm$0.02) and for family D is (R\textsubscript{D}(70$\degree$)=0.69$\pm$0.02), of the same order of what is expected from the simulation: 0.625. This is more than twice the value achievable when all NV centers contribute equally to the total PL.
Concerning sensitivity (\textrho\textsubscript{i}~and \textchi\textsubscript{i}~parameters, \cref{fig:fig5}g-h), also in this case it behaves similarly to contrast. The maxima of  \textrho\textsubscript{B} (\textrho\textsubscript{B}(15\degree)=10.8$\pm$0.3) and \textrho\textsubscript{D} (\textrho\textsubscript{D}(70\degree)=16.3$\pm$0.5) are respectively 2.2 and 2.8 times larger than the value of \textrho\textsubscript{B}(0$\degree$) and \textrho\textsubscript{D}(0$\degree$) in \cref{fig:fig5}c, that is when the four NV center families contribute equally to the PL detected. This improvement is in agreement with the value estimated from the simulation (ratio between the maximum of \textchi\textsubscript{i} in \cref{fig:fig5}h and \textchi\textsubscript{i}(45$\degree$) in \cref{fig:fig5}d), which is 2.4.  

\section*{Conclusion}
In this paper, we investigated the contrast and the sensitivity dependence of an ensemble of NV centers when both the laser polarization and the polarizer orientation are tuned. Modelling and simulating the NV center excitation and emission processes, we showed the opportunity to improve the contrast and the sensitivity respectively of a factor of 2.5 and 2.4 as compared to the configuration in which all NV centers equally contribute to the total PL. We experimentally demonstrated a relative ODMR contrast of 60\% for a single NV center family and the opportunity to double its sensitivity by simply equipping the experimental set-up of a polarizer and a half-wave plate. Moreover, we showed how this technique allows identifying which NV family corresponds to each ODMR peak, which is useful for vector magnetometry applications or quantitative analysis of NV center preferential orientations distribution. Finally, applying this technique to diamond samples grown along $\langle$110$\rangle$ or $\langle$113$\rangle$, exhibiting NV centers with partial preferential orientation, may allow a further gain in sensitivity \cite{Pham2012} equivalent to what would be expected from high quality bulk diamond crystals showing 100\% of NV preferential orientation.

\appendix
\section*{Methods}

\subsection{NV center emission process}
\label{App:B}
The NV center PL emission process relies on the emission of the two NV center incoherent dipoles (\textbf{d\textsubscript{1}},\textbf{d\textsubscript{2}}). The total PL is the sum of the PL intensity emitted by each dipole \cite{Alegre2007} and reads as:
\begin{equation}
I=I_{d_1}+I_{d_2}
\end{equation}
where I\textsubscript{d\textsubscript{i}} is the PL intensity emitted by \textbf{d\textsubscript{i}}:
\begin{equation}
I_{d_i}\propto |\vectorbold{E_{d_i}}|^2\propto|\vectorbold{(u\cross n_{d_i})\cross u}|^2
\label{eq:dipole}
\end{equation}
\textbf{E\textsubscript{d\textsubscript{i}}} is the far field approximation of the electric field emitted by the dipole \textbf{d\textsubscript{i}} in the direction defined by the unitary vector \textbf{u} \cite{Ohtsu2004}; \textbf{n\textsubscript{d\textsubscript{i}}} is a unitary vector oriented along \textbf{d\textsubscript{i}}. The far field approximation is generally valid for experiments with ensemble of NV centers in bulk diamonds. On the contrary, depending on the experimental configuration, the near-field approximation may be required when working with nanodiamonds \cite{Christinck2020}.

The radiation pattern of an oscillating dipole (\cref{eq:dipole}) is not spherical and both the intensity and the polarization of the emitted radiation depend on its propagation direction.
For simplicity, we assume that only the PL emitted along the optical axis of the imaging system is collected by the detector. In this way, using the notation of \cref{eq:dipole}, \textbf{u} corresponds to the optical axis of the imaging system and the axis of the polarizer in the detection system (\textbf{n\textsubscript{P}}) is perpendicular to \textbf{u}.  
This approximation is justified by the high diamond refractive index ($n=2.4$). In fact, using a microscope objective (NA=0.28) to collect the PL from a planar diamond facet perpendicular to the optical axis of the imaging system and placed in air ($n_2=1$), the solid angle over which the PL is collected is characterized by the angle ($\theta_2$) given by:
\begin{equation}
\theta_2=\arcsin(NA/n_1)=7\degree
\end{equation}

Under this approximation, and using the equality:
\begin{equation}
\vectorbold{(a\cross b)\cross c}=\vectorbold{(a\cdot c)b}-\vectorbold{(b\cdot c)a}
\end{equation}
we can write the PL component emitted by the dipole d\textsubscript{i} and polarized along \textbf{n\textsubscript{P}} (\textbf{n\textsubscript{P}} $\perp$ \textbf{u}) as:
\begin{equation}
I_{d_i}(\vectorbold{n_P})\propto |\vectorbold{E_{d_i}\cdot n_P}|^2\propto|(\vectorbold{(u\cross n_{d_i})\cross u})\cdot \vectorbold{n_P}|^2=|\vectorbold{(u\cdot u)(n_{d_i}\cdot n_P)-(n_{d_i}\cdot u)(u\cdot n_P)}|^2=|\vectorbold{n_{d_i}\cdot n_P}|^2
\end{equation}
Therefore the total PL component (I(\textbf{n\textsubscript{P}})) polarized along \textbf{n\textsubscript{P}} reads as:
\begin{equation}
I(\vectorbold{n_{P}})\propto |\vectorbold{n_{d_2}\cdot n_P}|^2+|\vectorbold{n_{d_1}\cdot n_P}|^2=1-|\vectorbold{n_{NV}\cdot n_{P}}|^2
\end{equation}
(\textbf{n\textsubscript{NV}},\textbf{n\textsubscript{d1}},\textbf{n\textsubscript{d2}}) is a set of three orthonormal vectors oriented respectively along the NV center axis and the two, arbitrary oriented, NV centers dipoles.
\subsection{Laser Excitation and PL collection along a $\langle100\rangle$ direction}
\label{App:C}
According to the diamond tetrahedral structure, the angle between a $\langle111\rangle$ and a $\langle100\rangle$ direction is $\acos(1/\sqrt{3})$. Therefore, when the electric field (\textbf{n\textsubscript{L}}) and the PL polarization (\textbf{n\textsubscript{P}}) are along a $\langle100\rangle$ direction, it results: 
\begin{equation}
P_i(\vectorbold{n_{L}})=I_i(\vectorbold{n_{P}})=1-\abs{\frac{1}{\sqrt{3}}}^2=\frac{2}{3}
\end{equation} 
Consequently, we obtain:
\begin{equation}
\chi_i(\vectorbold{n_{L}},\vectorbold{n_{P}})=\frac{P_i I_i}{\sqrt{\sum_{j=1}^4 P_j I_j}}=\frac{1}{3}
\end{equation} 
\subsection{Fitting procedure}
\label{App:D}
The ODMR spectra are normalized and fitted by the function:
\begin{equation}
f(\nu)=1-\sum_{i=1}^8c_i\frac{(a_i/2)^2}{(a_i/2)^2+(\nu-\nu_i )^2} 
\label{eq:fitfunction}
\end{equation}
Where \textnu\textsubscript{i},a\textsubscript{i},c\textsubscript{i}, are respectively the resonance frequency, the ODMR linewidth and the ODMR contrast of the i\textsuperscript{th} ODMR peak (see supplementary note 2).

\subsection{Identification between ODMR peaks and NV center families}
\label{App:E}
\begin{figure}
	\centering
	\includegraphics[width=0.7\linewidth]{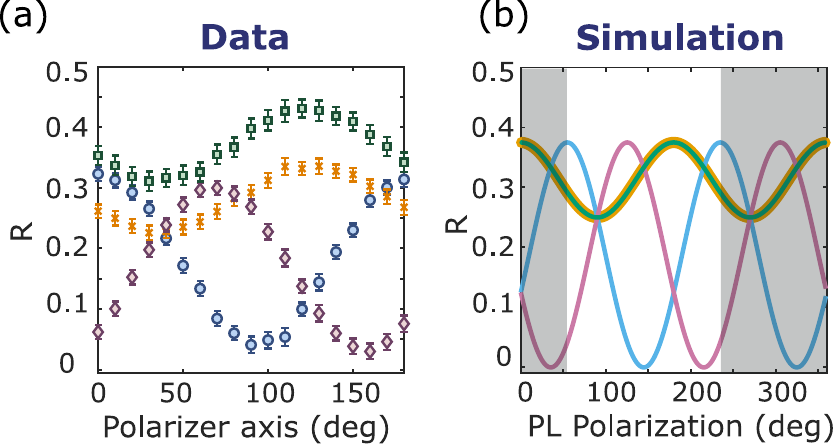}
	\caption{Measurement (a) and simulation (b) of the relative contrast of the four NV center families for different angles of the polarizer when the PL is collected from the top \{110\} diamond facet and the laser field is polarized along a $\langle100\rangle$ direction, as for \cref{fig:fig5}a-b. The color code is the same as in \cref{fig:fig4}. For the experimental results, the zero of the polarizer axis is arbitrarily chosen. For the simulation, it corresponds to the x axis of the laboratory reference frame (\cref{fig:fig4}b). Shadow areas are added a posteriori to match the measurement with the simulation.}
	\label{fig:SupFig2}
\end{figure}

In the main text we showed that tuning the laser polarization in a $\{110\}$ plane it is possible to completely suppress the PL contribution of one NV center family to the total PL aligning the laser field along the NV center axis of that family. In this way, in the main text, we identified the orientation of families B and D. A similar procedure has been performed collecting the PL from the top $\{110\}$ diamond facet and tuning the polarizer in this plane (\cref{fig:SupFig2}). That allowed suppressing, and thus identifying, the orientation of family A and C, the two families laying on the top $\{110\}$ plane.

\begin{figure}
	\centering
	\includegraphics[width=\linewidth]{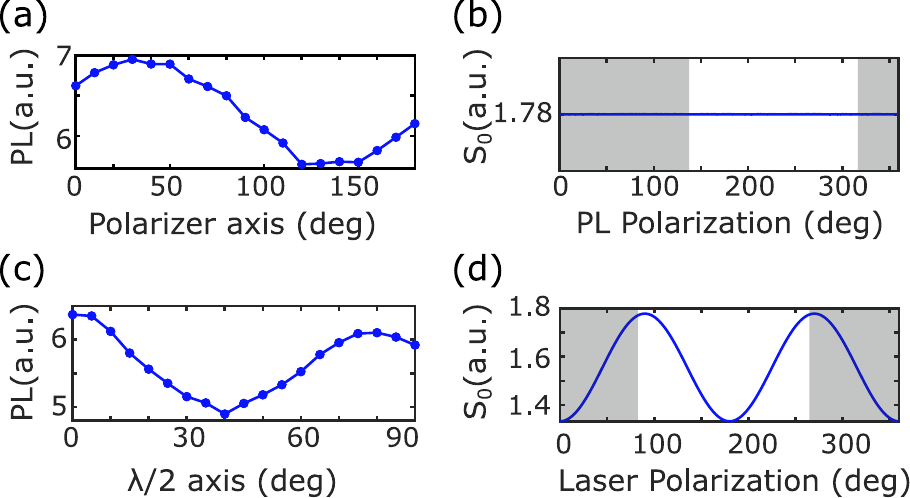}
	\caption{(a) Detected PL for different orientations of the polarizer axis. (b) Simulation of S\textsubscript{0} (\cref{eq:S0Polarization}) for different orientations of the PL polarization. (c) Detected PL for different orientations of the \textlambda/2 axis. (d) Simulation of S\textsubscript{0} (\cref{eq:S0Polarization}) for different orientations of the laser polarization.}
	\label{fig:SupFig1}
\end{figure} 

\subsection{Dependence of the detected PL on the polarizer angle and the laser polarization}
\label{App:F}
For a given laser polarization and polarizer orientation, the detected PL is defined as the mean value of the PL collected by the camera over 50 non-resonant frequency points. Its dependence on the orientation of the polarizer axis and \textlambda/2 axis is plotted respectively in \cref{fig:SupFig1}a and \cref{fig:SupFig1}c, and compared to the corresponding simulations in \cref{fig:SupFig1}b and \cref{fig:SupFig1}d. The simulation is realized using \cref{eq:S0Polarization} and setting $\alpha\frac{N\cdot \text{PL}_0^{\text{OFF}}}{4}=1$.
Concerning \cref{fig:SupFig1}a and \cref{fig:SupFig1}b, the electric field of the laser is oriented along a $\langle100\rangle$ axis and equally excites the four NV center families. The polarizer axis is rotated in a \{100\} plane. Although the simulation does not predict any PL variation, we measure small fluctuations ($\approx 20\%$) due to both experimental imperfections (e.g. aberrations caused by the rotation of the polarizer) and small differences in the PL emission rate (PL\textsubscript{0}\textsuperscript{ON/OFF}  \cref{eq:S0Polarization}) of the different families due to their different orientations with respect to the static magnetic field \cite{Tetienne2012}. 
Concerning \cref{fig:SupFig1}c and \cref{fig:SupFig1}d, the laser polarization is rotated in a \{110\} plane and we observe, both for data and simulation, PL fluctuations of around 25\%.
To conclude, PL variations when tuning the laser polarization and polarizer orientation are small compared to the corresponding ODMR contrast variations (\cref{fig:fig5}).

\section*{Data Availability}
The data that support the findings of this study are available from the corresponding author upon reasonable request.


\section*{Acknowledgements}
This project has received funding from the European Union’s Horizon 2020 research and innovation programme under grant agreement No. 820394 (ASTERIQS), the European Union’s Horizon Europe research and innovation program under grant agreement No. 101080136 (AMADEUS), the Marie Skłodowska-Curie grant agreement No. 765267 (QuSCo), the QUANTERA grant agreement ANR-18-QUAN-0008-02265 (MICROSENS), the EIC Pathfinder 2021 program under grant agreement No. 101046911 (QuMicro) and the EMPIR project 20IND05 (QADeT). We acknowledge the support of the Agence Innovation Défense under grant agreement ANR-22-ASM2-0003 (SPECTRAL).
The authors acknowledge Jean-François Roch and Philipp Neumann for fruitful discussions.

\section*{Author contributions}
S.M., L.M., X.L and T.D. contributed to the design and implementation of the research, to the analysis of the results and to the writing of the manuscript.

\section*{Additional Information}
All authors declare no competing interests.

\end{document}